\documentstyle[12pt]{article}
\begin{document}
\begin{flushright}
physics/0210117\\
SNBNCBS-2002
\end{flushright}
\vskip 2.7cm
\begin{center}
{\bf {\Large Jacobi Identity for Poisson Brackets: A Concise Proof}}

\vskip 3.5cm

{\bf R.P.Malik}
\footnote{ E-mail address: malik@boson.bose.res.in  }\\
{\it S. N. Bose National Centre for Basic Sciences,} \\
{\it Block-JD, Sector-III, Salt Lake, Calcutta-700 098, India} \\

\end{center}
\vskip 3.7cm

\noindent
{\bf Abstract:}               
In view of the recent interest in a short proof of the Jacobi identity
for the Poisson-brackets, we provide an alternative simple proof for the same. 
Our derivation is based on the validity of the Leibnitz rule in the context
of dynamical evolution.\\

\vskip 2cm

\noindent
{\it PACS number(s)}: 45.20.-d\\


\baselineskip=16pt


\newpage

\noindent
In the context of a precise {\it classical}
description of the particle dynamics, the Poisson brackets (PBs) play a 
very prominent role as far as the Hamiltonian formulation of the 
particle mechanics is concerned (see, e.g., [1,2]). The bilinearity, 
antisymmetry property and celebrated Jacobi identity, etc., are some of the 
key properties that are respected by PBs. Many of the text books on
classical mechanics provide the proof of Jacobi identity by exploiting the
tedious (but straightforward) algebra connected with the basic definition
of the PB in the (momentum) phase space [3]. ``There seems to be no simple 
way of proving Jacobi's identity for the Poisson bracket without lengthy 
algebra'' says Goldstein in his famous book [4]. To mitigate the complexity
of manipulations,  a certain specific set of differential operators are defined
in the well-known text books (see, e.g., [4, 5]). In a recent article [6],
a short proof for this identity has been given by using the concept of
infinitesimal canonical transformations for the dynamical variables. The
key role in this derivation is played by the {\it generator formalism} [4]
which owes its origin to the innate group property associated with the above 
infinitesimal transformations. The purpose of the present note is 
to demonstrate that there
is yet another simple proof for the Jacobi identity where (i) mainly the
Leibnitz rule is exploited for the derivation, and (ii)
the emphasis is laid on the dynamical evolution of the system. 
The Leibnitz rule plays a very
important and decisive 
role in the Hamiltonian description of the particle dynamics.
In fact, a dynamical system is said to be Hamiltonian [2] if and only if
the time derivative acts on a PB as if the latter
were a product of two dynamical
variables (see, e.g., (2) below).

Let $f (q,p)$ and
$ g (q,p) $ be a couple of dynamical variables, defined in the 
(momentum) phase space. It is evident that, right from the beginning,
 these variables have no explicit time dependence. Thus, their
time evolution w.r.t. the Hamiltonian function $H (q, p)$ is
$$
\begin{array}{lcl}
{\displaystyle \frac{ d f } { d t}} = 
{\displaystyle \frac{\partial f} {\partial q_{i}}}
\dot q_{i} + 
{\displaystyle \frac{\partial f} {\partial p_{i}}  \dot p_{i}} 
= {\displaystyle 
\frac {\partial f} {\partial q_{i}} \frac{\partial H} {\partial p_{i}}}  
- {\displaystyle 
\frac {\partial f} {\partial p_{i}} \frac{\partial H} {\partial q_{i}}}
\equiv \Bigl \{ f, H \Bigr \}_{PB}, \qquad
{\displaystyle \frac{ d g }{d t} } = \Bigl \{ g, H \Bigr \}_{PB},
\end{array}\eqno (1) 
$$
where $i = 1, 2, 3....s$ corresponds to the $s$ number of degrees of
freedom associated with the mechanical system. Here the summation convention
is adopted for the definition of the basic PB in the phase space
characterized by $(q_i, p_i)$. The application of the
Leibnitz rule leads to
$$
\begin{array}{lcl}
{\displaystyle \frac{ d  } { d t}} \Bigl \{ f, g \Bigr \}_{PB} =
\Bigl \{ {\displaystyle \frac{d f} { d t}}, g \Bigr \}_{PB}
+ \Bigl \{ f, {\displaystyle \frac{d g} {d t}}  \Bigr \}_{PB}.
\end{array}\eqno (2) 
$$
The above equation is valid {\it even} for the case when $f$ and $g$
are explicitly dependent on time (see, e.g., [5]). Now exploiting the
basic definition (1) of the time evolution for a dynamical variable, it
can be seen that each term of (2) can be separately expressed as 
$$
\begin{array}{lcl}
{\displaystyle \frac{ d  } { d t}} \Bigl \{ f, g \Bigr \}_{PB} &=&
\Bigl \{ \bigl \{ f , g \bigr \}_{PB}, H \Bigr \}_{PB}, \qquad
\Bigl \{ {\displaystyle \frac{d f} { d t}}, g \Bigr \}_{PB}
= \Bigl \{ \bigl \{ f, H \bigr \}_{PB}, g \Bigr \}_{PB}, \nonumber\\
 \Bigl \{ f, {\displaystyle \frac{d g} {d t}}  \Bigr \}_{PB} 
&=&  \Bigl \{ f, \bigl \{ g , H \bigr \}_{PB} \Bigr \}_{PB}.
\end{array}\eqno (3) 
$$
Substitutions of the above expressions into (2) and rearrangements 
(corresponding mainly to the antisymmetry property of the PB)
yield the Jacobi identity for all the above three 
dynamical variables as
$$
\begin{array}{lcl}
\Bigl \{ \bigl \{ f, g \bigr \}_{PB}, H \Bigr \}_{PB} +
\Bigl \{ \bigl \{ g, H \bigr \}_{PB}, f \Bigr \}_{PB} +
\Bigl \{ \bigl \{ H, f \bigr \}_{PB}, g \Bigr \}_{PB} = 0.
\end{array}\eqno (4)
$$
As far as the emphasis on the evolution of the dynamical variable is concerned, 
the derivation for the Jacobi identity in this note 
highlights the key role played by the Leibnitz rule in the context of 
Hamiltonian description of the particle mechanics. Of course, here the
Hamiltonian $H$ plays similar kind of exceptional role as the generator $h$ 
plays for the infinitesimal canonical transformation $\delta_{c}$. For the 
derivation of Jacobi identity in [6], one starts with the application
of $\delta_{c}$ on a PB as 
$$
\begin{array}{lcl}
\delta_{c} \bigl \{ f, g \bigr \}_{PB} =
\Bigl \{ \bigl \{ f, g \bigr \}_{PB}, h \Bigr \}_{PB}. 
\end{array}\eqno (5)
$$
The l.h.s. of the above equation can be further expanded as
$ \{ \delta_{c} f, g \}_{PB} + \{ f, \delta_{c} g \}_{PB}$. Now, exploiting
the basic definition of the generator $h$ (i.e. $\delta_{c} f 
= \{ f, h \}_{PB} $) and making some rearrangement 
in the above equation, one derives the Jacobi identity as [6]
$$
\begin{array}{lcl}
\Bigl \{ \bigl \{ f, g \bigr \}_{PB}, h \Bigr \}_{PB} +
\Bigl \{ \bigl \{ g, h \bigr \}_{PB}, f \Bigr \}_{PB} +
\Bigl \{ \bigl \{ h, f \bigr \}_{PB}, g \Bigr \}_{PB} = 0.
\end{array}\eqno (6)
$$
It should be noted that the generator $H$ for the dynamical evolution
and the generator $h$ for the infinitesimal canonical transformation
are on equal footing as far as the derivation of the Jacobi
identity in (4) and (6) is concerned. Whereas the emphasis in the 
former case is on the ``dynamical evolution'', the latter case revolves
around the concept of ``infinitesimal symmetry'' and its innate group
property.

\end{document}